\documentclass[
 reprint,
 amsmath,amssymb,
 aps
]{revtex4-2}

\usepackage{graphicx}
\usepackage{dcolumn}
\usepackage{bm}
\usepackage[normalem]{ulem}
\usepackage{xcolor}
\usepackage{newtxtext}
\usepackage{lmodern,pdftexcmds}
\usepackage{xstring}
\usepackage{mathrsfs}
\usepackage{soul}

\DeclareMathAlphabet{\mathbfsf}{\encodingdefault}{\sfdefault}{bx}{sl}
\renewcommand{\Vec}[1]{\bm{#1}}
\def\Tens#1{\IfSubStr{ABCDEFGHIJKLMNOPQRSTUVWXYZabcdefghijklmnopqrstuvwxyz}{#1}{\mathbfsf{#1}}{\bm{#1}}}
\newcommand{\figref}[1]{Fig.\,\ref{#1}}
\newcommand{\Figref}[1]{Figure\,\ref{#1}}

\newcommand{\figrefp}[2]{Fig.\,\ref{#1}\,(#2)}
\newcommand{\Figrefp}[2]{Figure\,\ref{#1}\,(#2)}

\begin{document}

\preprint{APS/123-QED}

\title{Odd Viscosity in Chiral Passive Suspensions}

\author{Zhiyuan \surname{Zhao}%
$^{1,2}$}%
\author{Mingcheng \surname{Yang}$^{2,3,4}$}%
\author{Shigeyuki \surname{Komura}$^{1,5,6}$}%
\author{Ryohei \surname{Seto}$^{1,5,7}$}%
\email{seto@ucas.ac.cn}%
\affiliation{%
$^1$Wenzhou Institute, University of Chinese Academy of Sciences, Wenzhou, Zhejiang 325000, China}%
\affiliation{%
$^2$School of Physical Sciences, University of Chinese Academy of Sciences, Beijing 100049, China}
\affiliation{%
$^3$Beijing National Laboratory for Condensed Matter Physics and Laboratory of Soft Matter Physics, Institute of Physics,
Chinese Academy of Sciences, Beijing 100190, China}%
\affiliation{%
$^4$Songshan Lake Materials Laboratory, Dongguan, Guangdong 523808, China}%
\affiliation{%
$^5$Oujiang Laboratory (Zhejiang Lab for Rengerative Medicine, Vision and Brain Health), Wenzhou, Zhejiang 325000, China}%
\affiliation{%
$^6$Songshan Department of Chemistry, Graduate School of Science, 
Tokyo Metropolitan University, Hachioji, Tokyo 192-0397, Japan}
\affiliation{%
$^7$Graduate School of Information Science, University of Hyogo, Kobe, Hyogo 650-0047, Japan}

\date{\today}

%%%%%%%%%%%%%%%%%%%%%%%%%%%%%%%%%%%%%%%%%%%%%%%%%%%%%%%%%%%%%%%%%%%%%%%%%%%%
%%%%%%%%%%%%%%%%%%%%%%%%%%%%%%%%%%%%%%%%%%%%%%%%%%%%%%%%%%%%%%%%%%%%%%%%%%%%

\begin{abstract}
Prior studies have revealed that nonzero odd viscosity is an essential property for chiral active fluids. 
Here we report that such an odd viscosity also exists in suspensions of non-active or non-externally-driven but chirally-shaped particles.
Computational simulations are carried out for 
monolayers of dense ratchets 
in simple shear and planar extensional flows. 
The contact between two ratchets can be either frictionless or infinitely-frictional,
depending on their teeth and sliding directions at the contact point.
Our results show that the ratchet suspension has the intermediate shear/extensional viscosity 
as compared with the suspensions of smooth and gear-like particles.
Meanwhile, the ratchet suspensions show nonzero even and odd components of the first normal stress coefficient, 
which indicates the mixed feature of conventional complex fluids and chiral viscous fluids. 
\end{abstract}

\maketitle

%%%%%%%%%%%%%%%%%%%%%%%%%%%%%%%%%%%%%%%%%%%%%%%%%%%%%%%%%%%%%%%%%%%%%%%%%%%%
%%%%%%%%%%%%%%%%%%%%%%%%%%%%%%%%%%%%%%%%%%%%%%%%%%%%%%%%%%%%%%%%%%%%%%%%%%%%

\section{Introduction}
Chiral active fluids are typical nonequilibrium systems consisting of self-spinning constituents~\citep{tsai2005,soni2019,scholz2018}.
In recent years, they have attracted increasing attention because of intriguing dynamics and collective behaviors
such as turbulence~\citep{reeves2021}, phase separation~\citep{shen2020,massana2021,zhao2021}, surface wave~\citep{soni2019,abanov2018}, and unidirectional edge current~\citep{van2016,souslov2019,yang2020,liu2020,yang2021}. 
One essential property to explain such behaviors is the so-called Hall or odd viscosity,
which stems from the inherent breaking of parity and time-reversal symmetries 
and does not produce any entropy or heat as dissipative viscosity~\citep{avron1998,banerjee2017}. 
Prior work on the odd viscosity ranges from phenomenological to topological and rheological scopes.
In most of such work, the self-spinning constituents are sufficiently small,
so that the chiral active fluids are treated as continuum phases.
Then their behaviors can be described by hydrodynamic equations with an additional assumed odd term,
where the odd viscosity physically characterizes 
the orthogonal stress response of the system 
to the imposed flow (i.e., eigendirections of the rate of deformation tensor)%
~\citep{ganeshan2017,liao2019,hargus2020,epstein2020,han2021,hosaka2021,hosaka2021hydrodynamic}.

In the field of rheology, 
the first normal stress difference,
as one of viscometric functions,
is widely studied~\citep{Coleman_1966,bird1987dynamics,dbouk2013}.
It also describes the orthogonal stress response to the imposed flow,
but is commonly discussed for conventional complex fluids such as viscoelastic fluids and dense suspensions.
We have clarified the relation between the odd viscosity and the first normal stress difference in the recent work~\citep{zhao2021}.
In general, the latter can be decomposed into even and odd components.
The even one results from the microstructures constructed by interacting fluid constituents,
whereas the odd one corresponds to the odd viscosity when the parity and time-reversal symmetries of the system break. 
Our work reported that both of such even and odd components were nonzero for chiral active suspensions,
which indicated the mixed feature of conventional complex fluids and chiral viscous fluids~\citep{zhao2021}. 
Nevertheless, 
the understanding of rheology of chiral active suspensions is still lacking.
Another open question is whether the odd viscosity only exists in active fluids
(with self-spinning or externally-driven constituents)
or not.

In this paper, we examine the odd viscosity in passive suspensions composed of chiral particles with finite size,
by carrying out computational simulations in both simple shear and planar extensional flows \citep{Seto_2017}. 
The chiral particles are modeled by ratchets with unidirectional (clockwise or anticlockwise) teeth,
which are supposed to undergo asymmetric contact interactions.
This means the contact can be either frictional or frictionless depending on the teeth and sliding directions of the particles. 
In section 2, we detail the main simulation methods, 
including the modeling of particles dynamics and background flows, simulation parameters and conditions, 
and rheological characterization. 
Section 3 presents the simulation results 
in terms of average contact number of the particles,
reorientation angles,
and even and odd components of essential rheological coefficients.

%%%%%%%%%%%%%%%%%%%%%%%%%%%%%%%%%%%%%%%%%%%%%%%%%%%%%%%%%%%%%%%%%%%%%%%%%%%%
%%%%%%%%%%%%%%%%%%%%%%%%%%%%%%%%%%%%%%%%%%%%%%%%%%%%%%%%%%%%%%%%%%%%%%%%%%%%

\section{Simulation Method}

\subsection{Particle dynamics}
For $N$ spherical particles that are suspended in liquid solvent,
they experience forces and torques due to 
Stokes drag ($\Vec{F}_{\mathrm{S}}$ and $\Vec{T}_{\mathrm{S}}$),
hydrodynamic inter-particle interactions ($\Vec{F}_{\mathrm{H}}$ and $\Vec{T}_{\mathrm{H}}$),
and frictional contact ($\Vec{F}_{\mathrm{C}}$ and $\Vec{T}_{\mathrm{C}}$). 
When the flow time scale is shorter than the Brownian time scale, 
we can neglect both inertia and thermal fluctuations.
As a result, the force and torque balances on particle $i$ ($i=1,\dotsc,N$) are given by
\begin{gather}
  \Vec{F}_{\mathrm{S},i} 
  +
  \Vec{F}_{\mathrm{H},i} 
  + 
  \sum_{j \neq i}
  \Vec{F}_{\mathrm{C},ij}
  =   \Vec{0},
  \label{eq:force_balance} \\
  \Vec{T}_{\mathrm{S},i} 
  + 
  \Vec{T}_{\mathrm{H},i} 
  + 
  \sum_{j \neq i}
  \Vec{T}_{\mathrm{C},ij}
  =
  \Vec{0}.
  \label{eq:torque_balance}
\end{gather}
Here, the Stokes force and torque are given by
\begin{gather}
  \Vec{F}_{\mathrm{S},i}
  =
  -6 \pi \eta_0 a \left( \Vec{U}_i - \Vec{U}^{\infty} (\Vec{x}_i) \right), 
  \label{eq:stokes_force} \\
  \Vec{T}_{\mathrm{S},i}
  =
  -8 \pi \eta_0 a^3 \left( \Vec{\Omega}_i - \Vec{\Omega}^{\infty} (\Vec{x}_i) \right),
  \label{eq:stokes_torque}
\end{gather}
where $\eta_0$ represents the solvent viscosity, 
$a$ the particle radius,
$\Vec{U}_i$ and $\Vec{\Omega}_i$ the velocity and angular velocity of particle $i$, respectively,
and $\Vec{U}^{\infty} (\Vec{x}_i)$ and $\Vec{\Omega}^{\infty} (\Vec{x}_i)$ 
the velocity and angular velocity of the background fluid at particle position $\Vec{x}_i$, respectively.
For the inter-particle hydrodynamic interactions,
we assume that they only arise from lubrication effects. 
This is justified for dense suspensions subjected to contact forces,
where the far-field or many-body hydrodynamic interactions play minor roles.
The $6N$ lubrication force and torque vectors 
($\Vec{F}_{\mathrm{H}} \equiv \{ \Vec{F}_{\mathrm{H},1}, \dotsc, \Vec{F}_{\mathrm{H},N} \}$ 
and $\Vec{T}_{\mathrm{H}} \equiv \{ \Vec{T}_{\mathrm{H},1}, \dotsc, \Vec{T}_{\mathrm{H},N} \}$)
are coupled with the $6N$ particle velocity and angular velocity vectors 
($\Vec{U} \equiv \{ \Vec{U}_1, \dotsc, \Vec{U}_N \}$ 
and $\Vec{\Omega} \equiv \{ \Vec{\Omega}_1, \dotsc, \Vec{\Omega}_N \}$) in the form of 
\begin{equation}
  \begin{pmatrix}
    \Vec{F}_{\mathrm{H}}\\
    \Vec{T}_{\mathrm{H}}
  \end{pmatrix} 
  = 
-\Tens{R}_{\mathrm{L}}
\cdot
\begin{pmatrix}
\Vec{U}-
\Vec{U}^{\infty} (\Vec{x}) \\ 
\Vec{\Omega} -
\Vec{\Omega}^{\infty} (\Vec{x})
 \end{pmatrix}
+
\Tens{R}'_{\mathrm{L}}
 : \Tens{E}^{\infty},
\end{equation}
where $\Tens{R}_{\mathrm{L}}$ and $\Tens{R}^{\prime}_{\mathrm{L}}$ are the configuration-dependent resistance matrices 
for the hydrodynamic lubrication,
and $\Tens{E}^{\infty}$ denotes the rate-of-strain tensor~\citep{brady1988}.
In the current work,
the resistance matrices are simply described by the leading terms of the pairwise short-range lubrication interaction~\citep{ball1997}.

For two particles in contact, 
their interaction is described by a simple spring-and-dashpot model~\citep{luding2008,mari2014},
where the normal and tangential components of the contact force are given by 
\begin{gather}
    \Vec{F}^{(\mathrm{n})}_{\mathrm{C},ij}
    =
    k_{\mathrm{n}} h_{ij} \Vec{n}_{ij} 
    + 
    \gamma_{\mathrm{n}} \Vec{U}^{(\mathrm{n})}_{ij},
    \label{eq:contact_force_norm}
    \\
    \Vec{F}^{(\mathrm{t})}_{\mathrm{C},ij}
    =
    k_{\mathrm{t}} \Vec{\xi}_{ij}.
    \label{eq:contact_force_tan} 
\end{gather}
Here, $k_{\mathrm{n}}$ and $k_{\mathrm{t}}$ are the normal and tangential spring constants, respectively, 
$h_{ij}$ and $\Vec{n}_{ij}$ represent the surface separation and center-to-center unit vector between the particles, respectively,
$\gamma_{\mathrm{n}}$ is the damping constant,
$\Vec{U}^{(\mathrm{n})}_{ij} \equiv \Vec{n}_{ij} \Vec{n}_{ij} \cdot \left( \Vec{U}_{j} - \Vec{U}_{i} \right)$ is the relative normal velocity,
and $\Vec{\xi}_{ij}$ denotes the tangential stretch vector.
The contact forces fulfill Coulomb's friction law 
$|\Vec{F}^{(\mathrm{t})}_{\mathrm{C},ij}| \le \mu |\Vec{F}^{(\mathrm{n})}_{\mathrm{C},ij}|$ with the static friction coefficient $\mu$.
In addition, we note that $\Vec{\xi}_{ij} = \Vec{0}$ in the absence of contact ($h_{ij} > 0$).
After the particles contact at time $t_0$ ($h_{ij} \le 0$),
the tangential stretch vector evolves as $\Vec{\xi}_{ij}(t) =\int_{t_0}^{t} \Vec{U}^{\mathrm{(t)}}_{ij} dt$,
with the relative tangential velocity defined by
$\Vec{U}^{\mathrm{(t)}}_{ij} \equiv 
(\Tens{I} - \Vec{n}_{ij} \Vec{n}_{ij}) \cdot 
[\Vec{U}_j - \Vec{U}_i - (a \Vec{\Omega}_i + a \Vec{\Omega}_j) \times \Vec{n}_{ij}]$.
Then the tangential contact torque in Eq.~\eqref{eq:torque_balance} is obtained by
\begin{equation}
    \Vec{T}_{\mathrm{C},ij}
    =
    a \Vec{n}_{ij} \times \Vec{F}^{(\mathrm{t})}_{\mathrm{C},ij}.
    \label{eq:contact_torque}
\end{equation}

%%%%%%%%%%%%%%%%%%%%%%%%%%%%%%%%%%%%%%%%%%%%%%%%%%%%%%%%%%%%%%%%%%%%%%%%%%%%
%%%%%%%%%%%%%%%%%%%%%%%%%%%%%%%%%%%%%%%%%%%%%%%%%%%%%%%%%%%%%%%%%%%%%%%%%%%%

\begin{figure*}
\includegraphics[width=0.8\textwidth]{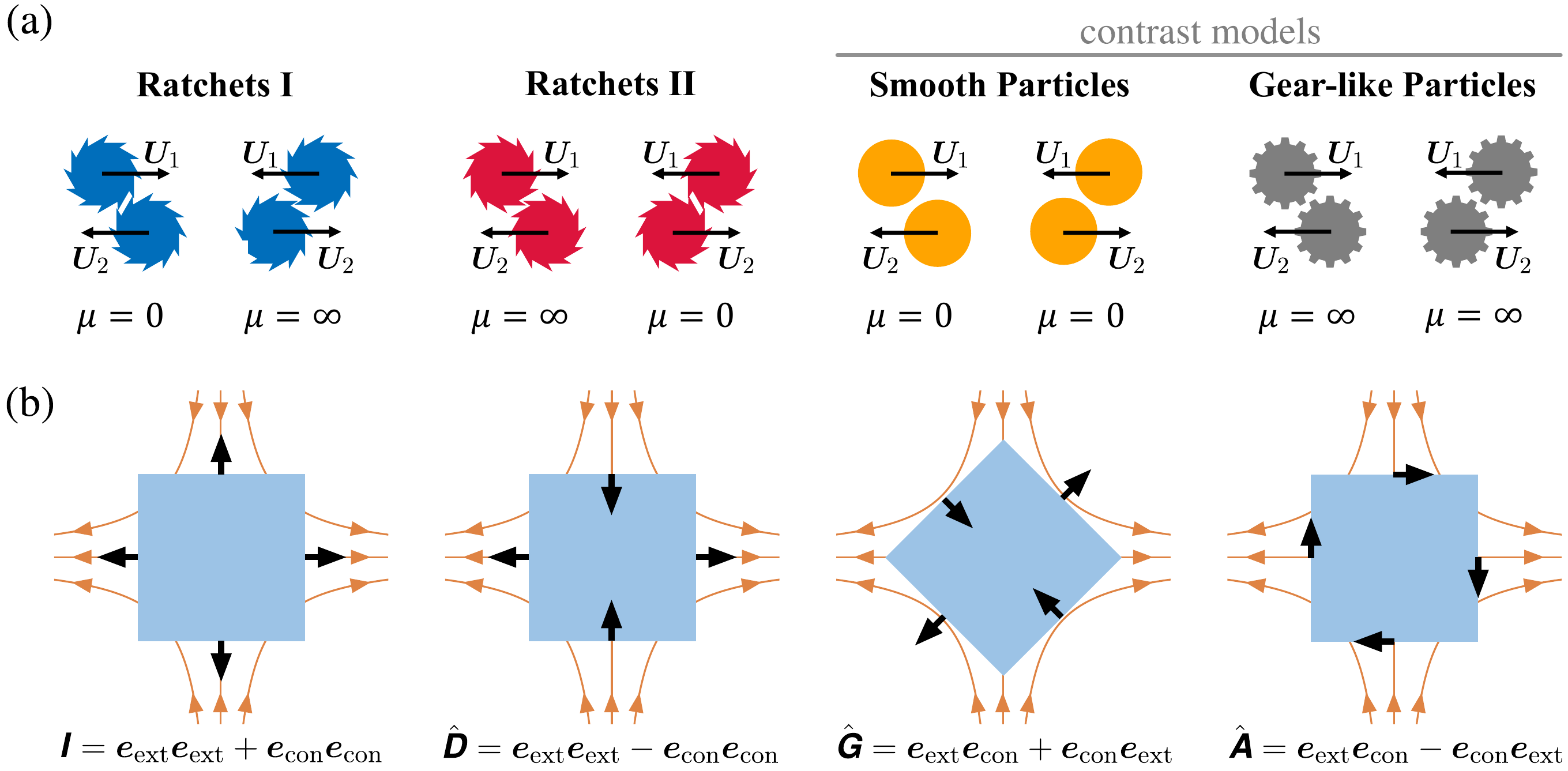}
\caption{
(a) Schematic of four particle models:
ratchets I (with clockwise teeth), ratchets II (with anticlockwise teeth), smooth particles, and gear-like particles.
When two ratchets tangentially contact and and their teeth at the contact point are anti-parallel and parallel to their relative velocity,
we assume their interactions are frictionless ($\mu = 0$) and infinitely-frictional ($\mu = \infty$), respectively.
The smooth and gear-like particles (for contrast only) experience zero and infinite friction, respectively.
(b) Schematic of stress components and corresponding basis tensors. 
The black arrows indicate the stress directions,
whereas the orange lines and arrows
represent the embedded extensional flow
including the contraction (vertical) and extension (horizontal) axes.}
\label{fig:1}
\end{figure*}

\begin{figure*}
\includegraphics[width=0.8\textwidth]{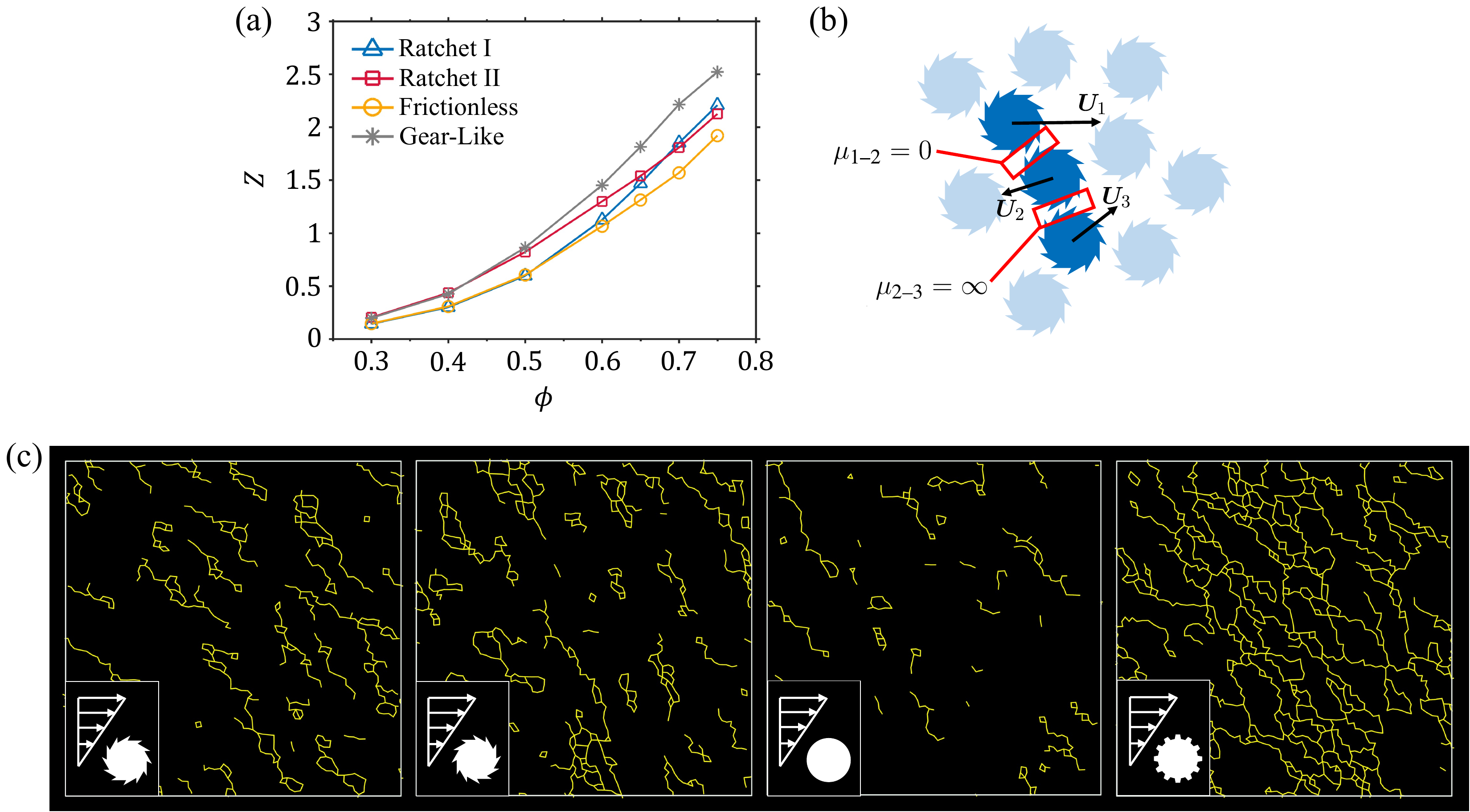}
\caption{
(a) Average contact number $Z$ as a function of particle areal fraction $\phi$ for various particle models.
Error bars are not shown because they are smaller than the symbols.
(b) Schematic of contact interaction between ratchets in the dense limit,
where frictionless and infinitely-frictional contacts emerge simultaneously. 
Here $U_1$, $U_2$, and $U_3$ denote the velocities of three contacted particles,
whereas $\mu_{1-2}$ and $\mu_{2-3}$ represent the static friction coefficients for particles 1 and 2 and particles 2 and 3, respectively.
(c) Representative snapshots of force chain distribution in simple shear flows
for particle areal fraction $\phi = 0.65$ and various particle models.
Inserts denote the particle type and the velocity gradient of the simple shear flow. }
\label{fig:2}
\end{figure*}

\subsection{Simulation parameters and conditions}
Our work takes into account four different types of particles, 
i.e., ratchet-like particles with clockwise teeth (ratchet I), ratchet-like particles with anticlockwise teeth (ratchet II), 
smooth particles, 
and gear-like rough particles (see \figrefp{fig:1}{a}).
For the smooth and gear-like particles, 
we assume their contacts are frictionless (i.e., $\mu = 0$) 
and infinitely-frictional (i.e., $\mu = \infty$), respectively. 
However, for ratchet-like particles, 
we assume the contact is frictionless/infinitely-frictional when a particle slides parallelly/anti-parallelly with respect to its tooth direction at the contact point.

Suspensions of the particles of the same type are exposed to constant simple shear and planar extensional flows,
which are constructed with the Lees–Edwards boundary condition~\citep{Lees1972}
and Kraynik–Reinelt periodic boundary condition~\citep{kraynik1992,Seto_2017}, respectively. 
The velocity field of the simple shear flow can be expressed as 
$\Vec{U}^{\infty}(\Vec{x}) = \Vec{\Omega}^{\infty}\times \Vec{x} +  \Tens{E}^{\infty} \cdot \Vec{x}$.
When the shear rate $\dot{\gamma}$ is constant,
we have the nonzero elements $U^{\infty}_{x} = \dot{\gamma}y$, 
$\Omega^{\infty}_{z} = -\dot{\gamma}/2$, 
and $E^{\infty}_{xy} = E^{\infty}_{yx} = \dot{\gamma}/2$. 
For the planar extensional flow, on the other hand, 
the velocity field is given by $\Vec{U}^{\infty}(\Vec{x}) = \Tens{E}^{\infty} \cdot \Vec{x}$ and
a constant extensional rate $\dot{\varepsilon}$ leads to the nonzero elements 
$U^{\infty}_{x} = \dot{\varepsilon}y$, $U^{\infty}_{y} = -\dot{\varepsilon}x$,  
and $E^{\infty}_{xy} = -E^{\infty}_{yx} = \dot{\varepsilon}$.

Simulations are carried out for $N = 3000$ bidisperse particles 
(with radii $a$ and $1.4a$ and with equal areal fractions) 
that are constrained in a monolayer ($x$-$y$ plane). 
The constants $\dot{\gamma}$ and $\dot{\varepsilon}$ are taken to be positive.
We set $k_\mathrm{n}$ and $k_\mathrm{t}$ (only for the cases of $\mu = \infty$) to sufficiently large values
that keep both the maximum overlap and tangential displacement smaller than $5 \%$ of the particle radius.
The particle areal fraction varies in the range of $0.3 \le \phi \le 0.75$.
For each set of simulation conditions and parameters, 
five parallel runs are performed starting from different random initial configurations.

%%%%%%%%%%%%%%%%%%%%%%%%%%%%%%%%%%%%%%%%%%%%%%%%%%%%%%%%%%%%%%%%%%%%%%%%%%%%
%%%%%%%%%%%%%%%%%%%%%%%%%%%%%%%%%%%%%%%%%%%%%%%%%%%%%%%%%%%%%%%%%%%%%%%%%%%%

\subsection{Rheological characterization}
The stress tensor of passive suspension can be obtained as 
\begin{equation}
    \Tens{\sigma} 
    = 
    2 \eta_0 \Tens{E}^{\infty}
    - \frac{1}{V} \sum_{i > j}
    \Vec{r}_{ij} 
    \left(
    \Vec{F}_{\mathrm{H},ij}
    +
    \Vec{F}_{\mathrm{C},ij}
    \right),
\end{equation}
where $V$ and $\Vec{r}_{ij}$ represent the total volume of the suspension
and center-to-center vector between particles $i$ and $j$, respectively. 
According to the theoretical framework discussed in reference~\cite{giusteri2018}, 
the stress tensor in two-dimensional systems can be decomposed in terms of basis tensors as 
\begin{equation}
    \Tens{\sigma} 
    = 
    -p \Tens{I} 
    + \dot{s} \big(
    \eta \hat{\Tens{D}}
    + \lambda \hat{\Tens{G}}
    + \zeta \hat{\Tens{A}}
    \big),
\end{equation}
where $p$ represents the pressure 
(including the isotropic stress due to contact forces), 
$\Tens{I}$ the identity tensor, 
$\dot{s}$ ($= \dot{\gamma}$ or $2\dot{\varepsilon}$) the flow rate, 
and $\lambda$ and $\zeta$ the non-dissipative response function and rotational viscosity, respectively.
The basis tensors are defined 
as $\hat{\Tens{D}} \equiv 
\Vec{e}_{\mathrm{ext}} \Vec{e}_{\mathrm{ext}} - \Vec{e}_{\mathrm{con}} \Vec{e}_{\mathrm{con}}$,
$\hat{\Tens{G}} \equiv \Vec{e}_{\mathrm{ext}} \Vec{e}_{\mathrm{con}} 
+ \Vec{e}_{\mathrm{con}} \Vec{e}_{\mathrm{ext}}$,
and $\hat{\Tens{A}} \equiv \Vec{e}_{\mathrm{ext}} 
\Vec{e}_{\mathrm{con}} - \Vec{e}_{\mathrm{con}} \Vec{e}_{\mathrm{ext}}$,
where $\Vec{e}_{\mathrm{ext}}$ and $\Vec{e}_{\mathrm{con}}$ 
are the unit vectors for 
the extension and contraction axes of the imposed flow.
We not that the basis tensors are orthogonal to each other and their corresponding stress components are shown in \figrefp{fig:1}{b}. 
%
% We note that $p_0$, $\eta$, $\lambda$, and $\zeta$ are functions of the flow type.
%
Besides, for suspensions without self-spinning elements, 
as in the current work,
the term with $\zeta$ can be dropped.

\begin{figure*}
\includegraphics[width=0.7\textwidth]{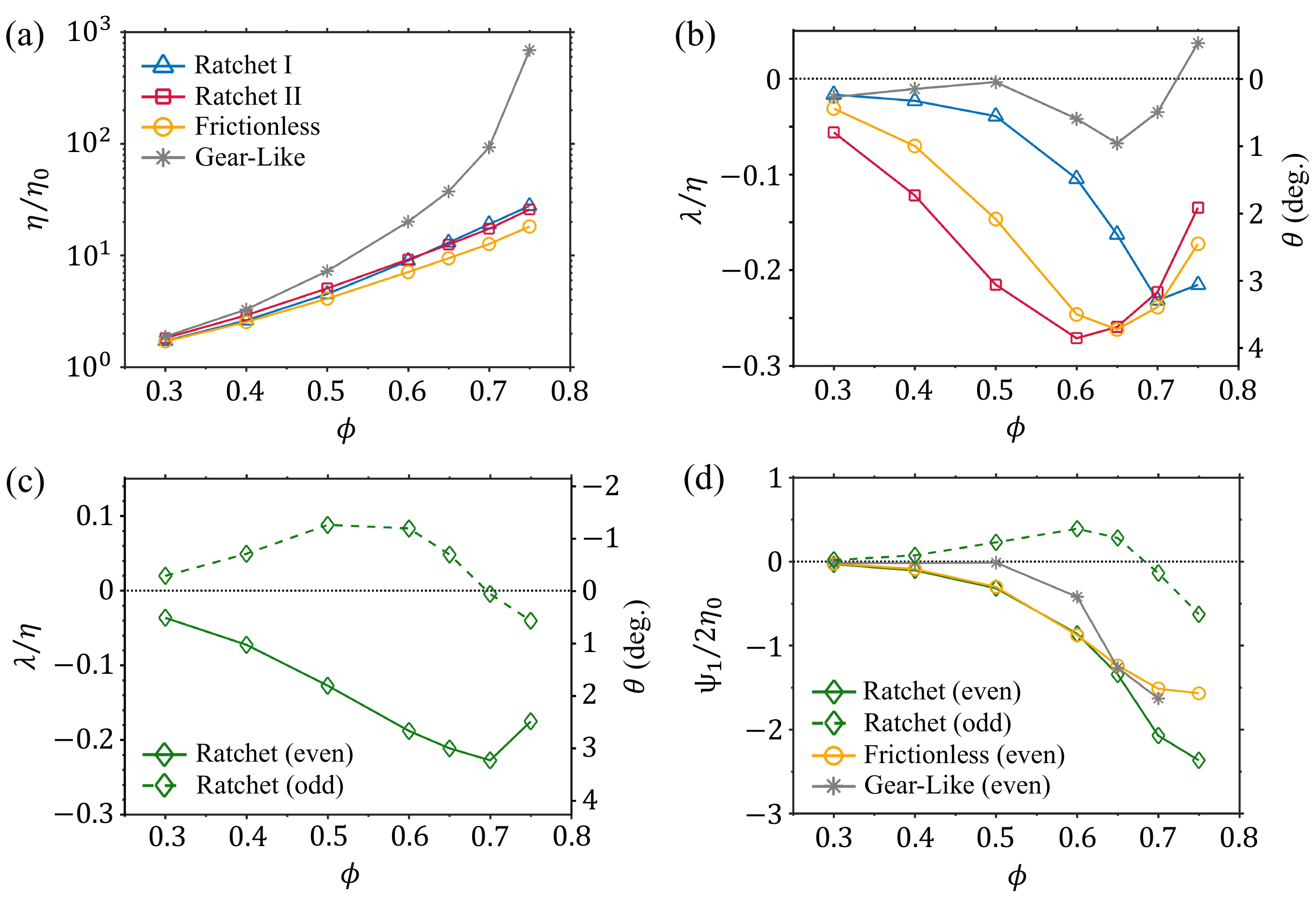}
\caption{
(a) Relative shear viscosity $\eta / \eta_0$ 
and (b) ratio of non-dissipation response function to shear viscosity $\lambda / \eta$ and corresponding reorientation angle $\theta$
as a function of particle areal fraction $\phi$ for various particle models.
(c) Even and odd components of the ratio $\lambda / \eta$ as a function of $\phi$ for ratchet particles.
(d) Scaled first normal stress coefficient as a function of $\phi$ for various particle models.
In the figures, error bars are not shown because they are smaller than the symbols. }
\label{fig:3}
\end{figure*}

Then the rotation of principal axes of $\Tens{\sigma}$ in the flow plane 
with respect to those of $\hat{\Tens{D}}$ is quantified by the reorientation angle
\begin{equation}
    \theta
    \equiv 
    \mathrm{arctan} \left[ \lambda / \left( \eta + \sqrt{\eta^2 +\lambda^2} \right) \right],
    \label{eq:reorientangle}
\end{equation}
which is proportional to the ratio $\lambda / \eta$, or equivalently, $N_1 / \sigma_{xy}$~\citep{giusteri2018}.
Here the first normal stress difference,
$N_1 \equiv \sigma_{xx} - \sigma_{yy}$,
is one typical signature indicating 
the presence of elasticity in complex fluids~\citep{seto2018,dbouk2013}.
It can alternatively be characterized through the first normal stress coefficient $\varPsi_1$ defined by 
\begin{equation}
    \Psi_1
    \equiv
    \frac{N_1}{|\dot{\gamma}|}
    =
    \frac{\dot{\gamma}}{|\dot{\gamma}|} \lambda.
    \label{eq:normalstressdifference}
\end{equation}
However, when planar extensional flows are imposed,
$\lambda$ does not depend on the sign of $\dot{\varepsilon}$
and the characterization in terms of $\Psi_1$ is unnecessary.
Therefore, we only study the non-dissipative response function $\lambda$ for the planar extensional flows. 
In order to further analyze the effect of shape chirality 
on any quantity $\varLambda$ of interest,
we decompose it into the even and odd components as
\begin{gather}
    \varLambda_{\mathrm{even}} 
    \equiv \frac{1}{2}
    \left[
    \varLambda(\dot{s}) + \varLambda(-\dot{s})
    \right],
    \label{eq:evencomponent}
    \\
    \varLambda_{\mathrm{odd}} 
    \equiv \frac{1}{2}
    \left[
    \varLambda(\dot{s}) - \varLambda(-\dot{s})
    \right].
    \label{eq:oddcomponent}
\end{gather}
When $\varLambda = \Psi_1 / 2$, 
the odd component $\varLambda_{\mathrm{odd}}$ corresponds to the odd viscosity~\citep{zhao2021}.

%%%%%%%%%%%%%%%%%%%%%%%%%%%%%%%%%%%%%%%%%%%%%%%%%%%%%%%%%%%%%%%%%%%%%%%%%%%%
%%%%%%%%%%%%%%%%%%%%%%%%%%%%%%%%%%%%%%%%%%%%%%%%%%%%%%%%%%%%%%%%%%%%%%%%%%%%

\section{Results}
For dense suspensions, 
both simple shear and planar extensional flows can give rise to particle contacts along the contraction axis.
The resultant particle force chains (or networks) and microstructures generate profound influences on the macroscopic rheology of the suspensions~\citep{larson1999,seto2013}.

Here, we first focus on the simulations of the simple shear flow
and estimate the average contact number $Z$ for various particle types and areal fractions (see \figrefp{fig:2}{a}).
For low areal fractions ($\phi \le 0.5$), 
it is observed that $Z < 1$ and the curves of ratchets I and II are in accordance with the curves of smooth and gear-like particles, respectively. 
Increasing the areal fraction leads the curves of ratchets I and II to coincide 
with intermediate $Z$ values as compared with those for the frictionless and gear-like particles.
Such a result is expected,
because, at low areal fractions, 
particles get into contact with one neighbor due to the applied shear flow.
The unique contact mode is referred to \figrefp{fig:1}{a} (left column for each particle type),
where the same friction coefficients are obtained between the ratchets I and smooth particles and between the ratchets II and gear-like particles.
However, for $\phi \ge 0.6$ the average contact number increases to $Z > 1$.
Being associated with the representative snapshots of inter-particle force chains, 
as shown in \figrefp{fig:2}{c},
we find that multi-particle contacts are dominant throughout the suspensions.
In this case, both the frictionless and infinitely-frictional contacts appear between the ratchet particles (see \figrefp{fig:2}{b}).
Especially for higher areal fractions,
the numbers of these contacts become similar.
Therefore, the behavior in \figrefp{fig:2}{a} is reasoned.

\begin{figure*}
\includegraphics[width=0.7\textwidth]{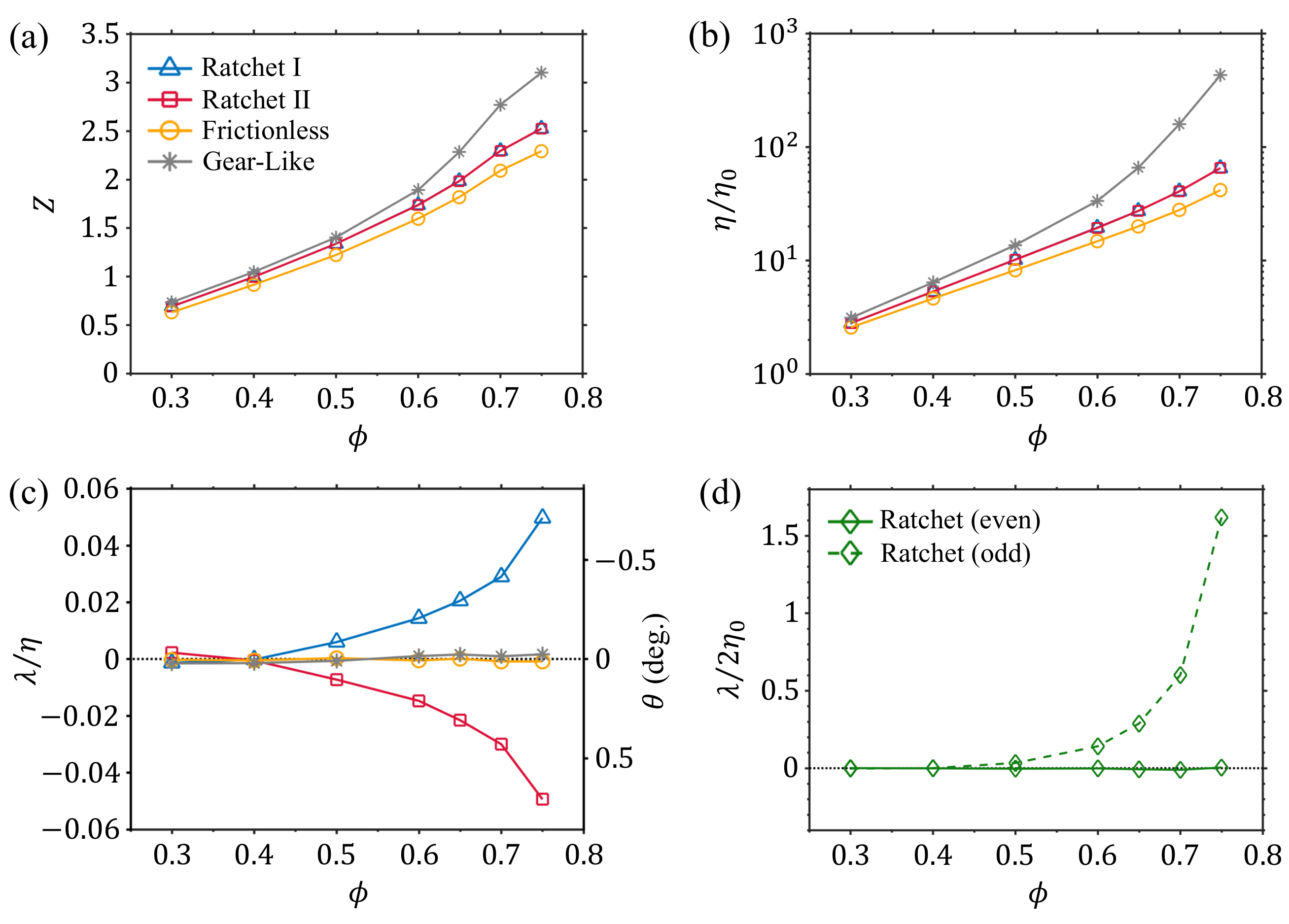}
\caption{
(a) Average contact number $Z$, 
(b) Relative extensional viscosity $\eta / \eta_0$, 
and (c) ratio of non-dissipation response function to extensional viscosity $\lambda / \eta$ and corresponding reorientation angle $\theta$
as a function of particle areal fraction $\phi$ for various particle models.
(d) Even and odd components of scaled non-dissipation response function $\lambda / (2 \eta_0)$ 
as a function of $\phi$ for ratchet particles.
In the figures, error bars are not shown because they are smaller than the symbols.}
\label{fig:4}
\end{figure*}

\Figrefp{fig:3}{a} shows the relative shear viscosity as a function of $\phi$ for different suspensions.
Because the inter-particle contact plays a decisive role,
the curves of relative shear viscosity show similar shapes with those presented in \figrefp{fig:2}{a}.
For the gear-like particles, 
the shear viscosity dramatically increases at $\phi = 0.75$ because of the proximity to the frictional jamming point. 
By comparing the relative shear viscosities between the ratchets and the other particles, 
we observe the former ones are in the intermediate values 
and still keep distance from the jamming state 
throughout the areal fraction studied.

\begin{figure*}
\includegraphics[width=0.75\textwidth]{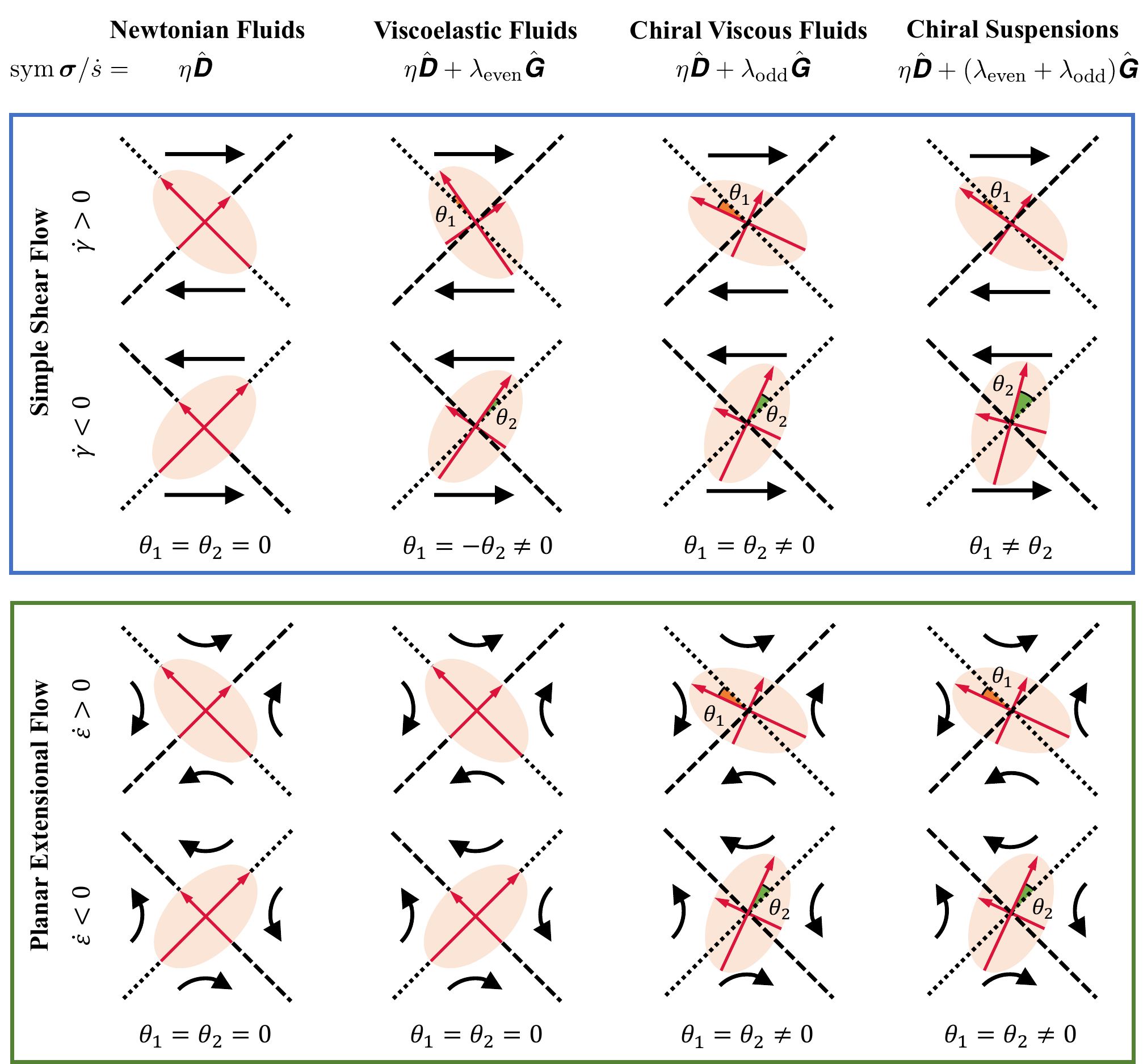}
\caption{
Schematics of response stress of different fluids in simple shear and planar extensional flows.
The red arrows represent the principal axes of the response stress in the flow plane,
whereas the black arrows denote the direction of imposed strain. 
The symbols $\lambda_{\mathrm{even}}$ and $\lambda_{\mathrm{odd}}$ 
denote the even and odd components of non-dissipative
response function $\lambda$.
The subscripts 1 and 2 indicate the reorientation angles for the positive and negative shear or extensional rate, respectively.}
\label{fig:5}
\end{figure*}

\Figrefp{fig:3}{b} shows the ratio of non-dissipative response function to shear viscosity $\lambda / \eta$, 
or equivalently, $N_1 / \sigma_{xy}$,
and the corresponding reorientation angle $\theta$ as a function of $\phi$ for various particle types. 
In the figure, 
the curves for the frictionless and infinitely-frictional particles agree with the results in the prior work~\citep{seto2018,royer2016}.
However, the curves for the ratchets I and II are not according, 
which indicates the essential role of particle chirality on determining the reorientation angle.
In order to give a further insight, 
we employ Eqs.~\eqref{eq:evencomponent} and \eqref{eq:oddcomponent} 
to decompose $\lambda / \eta$ into even and odd components 
for the ratchet particles.
As seen in \figrefp{fig:3}{c},
the even component takes negative values 
and its dependence on the areal fraction is typical of $\Psi_1$ for achiral particle suspensions \citep{Cwalina_2014}.
However, the finite odd component shows a completely different behavior, 
with positive values for $\phi \le 0.65$ and negative values for $\phi \ge 0.7$.
The vanishing odd component is near $\phi = 0.7$,
where the even component also experiences the turning point from decreasing to increasing.
Except $\phi = 0.7$, the odd component is non-negligible as compared with the even component.

In order to obtain the odd viscosity, 
we also calculate the even and odd components of 
the scaled first normal stress coefficient $\Psi_1 / (2 \eta_0)$. 
In \figrefp{fig:3}{d}, 
one can see that the ratchet particles have the similar even components with the frictionless particles for $\phi \le 0.6$, 
but smaller even components for $\phi \ge 0.65$.
The odd component for the ratchet particles 
shows the similar behavior with the odd component of $\lambda / \eta$ as shown in \figrefp{fig:3}{d}.
Such results demonstrate that both the features of conventional complex fluids and chiral viscous fluids 
exist in the passive ratchet suspensions.

In the following, we study the suspension rheology in the planar extensional flows. 
The average contact numbers $Z$ for different suspensions are shown in \figrefp{fig:4}{a}.
In the figure, we observe that throughout the areal fraction studied,
the $Z$ values for the ratchets I and II are similar
and intermediate as compared with
those for the frictionless and gear-like particles.
Such a phenomenon, 
not seen for the simple shear flows, 
is because the planar extensional flows are symmetric with respect to the contraction and extension axes.
Therefore, for both ratchets I or II, 
half of the particles undergo the frictionless contact and the other half experience the infinitely-frictional contact.
This explains the phenomena shown in \figrefp{fig:4}{b},
where extensional viscosities for the ratchets I and II also show the similar behaviors 
and the intermediate magnitudes with respect to those for the frictionless and gear-like particles.

Furthermore, since the first normal stress difference $N_1$ and coefficient $\Psi_1$ are introduced for simple shear flows, 
we cannot use them for planar extensional flows.
Thus,
we directly investigate the non-dissipative response function $\lambda$ instead. 
\Figref{fig:4}{c} shows the dependence of the ratio $\lambda / \eta$ and the corresponding reorientation angle on the particle areal fraction $\phi$.
In the figure, the values of $\theta$ for the frictionless and gear-like particles are almost zero, as expected.
Nevertheless, the ratchets I and II give rise to the monotonic increase and decrease, respectively, 
of the reorentation angle by increasing $\phi$.
By taking the decomposition of the scaled non-dissipative response function $\lambda / (2 \eta_0)$ for the ratchet particles, 
we observe the vanishing even component but prominent odd component that increases with $\phi$ (see \figrefp{fig:4}{d}). 
This result demonstrates that the reorientation angle in the planar extensional flows 
is only due to the existence of the odd viscosity.

%%%%%%%%%%%%%%%%%%%%%%%%%%%%%%%%%%%%%%%%%%%%%%%%%%%%%%%%%%%%%%%%%%%%%%%%%%%%
%%%%%%%%%%%%%%%%%%%%%%%%%%%%%%%%%%%%%%%%%%%%%%%%%%%%%%%%%%%%%%%%%%%%%%%%%%%%

\section{Discussions and conclusion}
Although the nonzero odd viscosity is obtained for chiral passive suspensions, 
we underline that its dependence on the particle areal fraction $\phi$ varies for different flow types. 
As exhibited in the result section, the simple shear and planar extensional flows 
give rise to the non-monotonic changes and monotonic increase of the odd viscosity, respectively.
This difference is due to the flow-induced microstructures,
which affect the non-dissipative response function $\lambda$ and 
then make the odd viscosity flow-type-dependent.
Besides, we demonstrate that the odd viscosity can also be characterized 
when the fluid constituents are not externally rotated (by active torques or imposed flows). 
Applying planar extensional flows is suggested to be 
a straightforward examination of 
the odd viscosity of a fluid.

In order to give a universal framework for the relevant rheological characterizations,
we schematically present in \figref{fig:5} the responsive stress and the corresponding reorientation angle for four different fluid systems.
Since the reorientation angle does not rely on the rotational stress response, 
our framework only considers the symmetric part of the stress
$\mathrm{sym}\,\Tens{\sigma}$.
Meanwhile, both the cases of simple shear and planar extensional flows are taken into account.

As seen in the first column, 
stable Newtonian fluids have a uniform distribution of constituents 
and thus constant viscosity.
The principal axes of $\mathrm{sym}\,\Tens{\sigma}$
should align exactly with those of $\hat{\Tens{D}}$.
However, for sheared conventional complex fluids (including viscoelastic fluids and dense suspensions),
the emergence of internal constituent microstructures 
leads the principal axes of 
$\mathrm{sym}\,\Tens{\sigma}$
to differ from those of $\hat{\Tens{D}}$.
Then one will obtain a nonzero response function $\lambda_{\mathrm{even}}$ 
(i.e., the even component of the first normal stress coefficient)
and the corresponding reorientation angles satisfying $\theta_1 = -\theta_2 \neq 0$.
We note that such behaviors are not typical for planar extensional flows,
because both of the flow field and resultant microstructures are symmetric 
with respect to the contraction and extension axes.
For chiral viscous fluids, 
the principal axes of the responsive stress rotate for both simple shear and planar extensional flows.
However, the origin is purely due to the intrinsic chirality, leading to
the loss of parity and time-reversal symmetries of the fluid.
As a result, the direction of the reorientation angle is independent of the shear or extensional rates, 
i.e., $\theta_1 = \theta_2 \neq 0$. 
The response function is the odd component of the first normal stress coefficient $\lambda_{\mathrm{odd}}$ or the odd viscosity.
Finally, in chiral suspensions (either active or passive), 
both microstructures and chirality contribute to the tilted principal axes of the responsive stress from those of $\hat{\Tens{D}}$.
Thus, the response function is the first normal stress coefficient 
(i.e., $\lambda_{\mathrm{even}} + \lambda_{\mathrm{odd}}$) for the simple shear flows, 
and is $\lambda_{\mathrm{odd}}$ for the planar extensional flows.

In conclusion, by carrying out computational simulations,
we have studies the rheology of passive chiral suspensions 
in constant simple shear and planar extensional flows.
The results of the shear and extensional viscosities show the intermediate values between those for the frictionless and gear-like particles. 
The dependence of the shear viscosity on the particle chirality 
is significant at low particle areal fractions but negligible for high areal fractions. 
Importantly, the chiral passive suspensions show nonzero even and odd components of the non-dissipative response function,
suggesting the mixed feature of conventional complex fluids and chiral viscous fluids. 
Such even and odd components have the comparable contributions to the reorientation angle of the system stress. 
We hope our work will extend the field of rheology and progress the understanding of chiral fluids.

%%%%%%%%%%%%%%%%%%%%%%%%%%%%%%%%%%%%%%%%%%%%%%%%%%%%%%%%%%%%%%%%%%%%%%%%%%%%
%%%%%%%%%%%%%%%%%%%%%%%%%%%%%%%%%%%%%%%%%%%%%%%%%%%%%%%%%%%%%%%%%%%%%%%%%%%%

\begin{acknowledgments}
Z.Z. acknowledge the Postdoctor Association of WIUCAS for helpful discussions. 
We also acknowledge Giulio Giusteri for valuable suggestions.
The work was supported by the startup fund of Wenzhou Institute, University of Chinese Academy of Sciences (No.WIUCASQD2020002 and No.WIUCASQD2021041),
National Nature Science Foundation of China (No.1217042129 and No.12174390),
and the Research Fund for International Scientists, National Nature Science Foundation of China (No.12150610463).
\end{acknowledgments}

%%%%%%%%%%%%%%%%%%%%%%%%%%%%%%%%%%%%%%%%%%%%%%%%%%%%%%%%%%%%%%%%%%%%%%%%%%%%
%%%%%%%%%%%%%%%%%%%%%%%%%%%%%%%%%%%%%%%%%%%%%%%%%%%%%%%%%%%%%%%%%%%%%%%%%%%%

\nocite{*}

\bibliography{library}

\end{document}